\documentclass[journal = Chemistry of Materials, manuscript = article, layout = traditional]{achemso}
\usepackage{natbib}
\usepackage{graphics,graphicx}
\usepackage[dvipsnames]{xcolor}
\usepackage{tabularx}
\usepackage[T1]{fontenc}
\usepackage{cancel}
\usepackage{times}
\usepackage{amssymb,amsfonts,amsmath}
\usepackage{oplotsymbl}
\usepackage{stix}
\usepackage{soul}

\usepackage[version=4]{mhchem}

\newcommand\edit[1]{\textcolor{black}{#1}} 

\usepackage{color}

\title{Predicted High-Pressure Hot Superconductivity in Li$_2$CaH$_{16}$ and Li$_2$CaH$_{17}$ Phases that Resemble the Type-II Clathrate Structure}

\author{Morgan Redington}
\affiliation{Department of Chemistry, State University of New York at Buffalo, Buffalo, NY 14260-3000, USA}
\author{Eva Zurek}
\email{ezurek@buffalo.edu}
\affiliation{Department of Chemistry, State University of New York at Buffalo, Buffalo, NY 14260-3000, USA}

\begin{document}

\newpage

\begin{abstract}
High-temperature high-pressure superconducting hydrides are typically characterized by cage-like hydrogenic lattices filled with electropositive metal atoms. Here, density functional theory based evolutionary crystal structure searches find two phases that possess these geometric features and are related to the Type-II clathrate structure. In these $Fd\overline{3}m$ Li$_2$CaH$_{16}$ and $R\overline{3}m$ Li$_2$CaH$_{17}$ phases the calcium atom occupies the larger cage and the lithium atom the smaller one. The highest superconducting critical temperatures predicted within the isotropic Eliashberg formalism, 330~K at 350~GPa for $Fd\overline{3}m$ Li$_2$CaH$_{16}$ and 370~K at 300~GPa for $R\overline{3}m$ Li$_2$CaH$_{17}$, suggest these structures are another example of hot superconducting hydrides. As pressure is lowered the cage-like lattices distort with the emergence of quasimolecular hydrogenic motifs; nonetheless Li$_2$CaH$_{17}$ is predicted to be superconducting down to 160~GPa at 205~K.
\end{abstract}



\maketitle


\newpage

\section{Introduction}
In 1968, Neil Ashcroft predicted that metallic hydrogen -- the hypothetical ground state of the lightest element under pressure -- could behave as a high-temperature superconductor~\cite{RN693}. \edit{The realization of this state has proven elusive, with the most recently estimated pressure needed being over 570~GPa~\cite{RN1076}}. Nearly four decades later Ashcroft described how the pressures required to metallize hydrogen could be decreased via addition of a second element that "chemically precompressed" the system~\cite{RN1}. As early as 2015, experimental studies supporting Ashcroft's prediction came to the fore. These included measurements of high superconducting critical temperatures ($T_\text{c}$s) in phases with unique stoichiometries, stable only under pressure, including H$_3$S ($T_\text{c}$ = 203~K, 150~GPa)~\cite{RN5}, LaH$_{10}$ ($T_\text{c}$ = 260~K, 200~GPa)~\cite{RN2,RN938}, YH$_{9}$ ($T_\text{c}$ = 262~K, 182~GPa)~\cite{RN921}, YH$_{6}$ ($T_\text{c}$ = 224~K, 166~GPa)~\cite{RN925}, CaH$_6$ ($T_\text{c}$ = 210-215~K, 160-172~GPa)~\cite{RN906,RN904,RN652}, and CeH$_9$~\cite{RN736,RN737,RN662} ($T_\text{c}$ = 100~K, 166~GPa). Recently, local magnetometry experiments on CeH$_9$, performed via nitrogen-vacancy centers embedded within the diamond anvil, provided evidence of the Meissner effect~\cite{RN1015}.

The immense pressures required to stabilize the superconducting binary hydrides can only be reached in diamond anvil cells and come coupled with experimental intricacy and expense, making theoretical calculations an appealing method to both determine systems of interest and interpret experimental results~\cite{RN43,RN1001}. Density functional theory (DFT) calculations \edit{have been found capable of predicting superconducting properties of conventional superconductors}, and crystal structure prediction (CSP) techniques have been successful in uncovering promising synthetic targets (for unit cells that are not too complex)~\cite{RN932}. The chemical space has been thoroughly scoured for fruitful high-temperature superconducting binary hydrides~\cite{RN732,RN623,RN750}. In contrast, ternary hydrides present relatively unexplored possibilities, however predictions for this class of hydrides are significantly more challenging~\cite{RN905}. The vast chemical space means that an increased number of stoichiometries must be considered for each elemental combination, with a more complex potential energy landscape that requires a greater number of structures be optimized in the CSP runs~\cite{RN910}. To overcome these challenges, machine learning potentials have been used to roughly explore the phase diagram prior to performing more costly DFT optimizations~\cite{RN1017}, and high-throughput calculations have been employed to study elemental substitutions on previously-discovered superconducting templates~\cite{RN1016}. The introduction of a third element could potentially be used to lower the stabilization pressure, such as in synthesized LaBeH$_8$ ($T_\text{c}$ = 110~K, 80~GPa)~\cite{PhysRevLett.130.266001} and predicted BaSiH$_8$ ($T_\text{c}$ = 71~K, 3~GPa)~\cite{RN1016}, or increase the superconducting critical temperature as in the synthesized (La,Ce)H$_9$ ($T_\text{c}$ = 148-178~K, 97-172~GPa)~\cite{RN1018} and predicted Li$_2$ScH$_{20}$ ($T_\text{c}$ = 242~K, 300~GPa)~\cite{PhysRevB.106.024519}. The former strategy is usually accomplished by using one electropositive element and one $p$-block element, whereas the latter strategy usually requires the combination of two electropositive elements. Recently, superconductivity in a quaternary hydride, (La,Y,Ce)H$_{10}$, has been reported with a maximum $T_\text{c}$ of 190~K at 112~GPa~\cite{Chen:2024}.

The hydrogenic lattices of the binary hydride superconductors of an electropositive element that have very high $T_c$s generally assume cage-like structures within which a metal atom is encapsulated. Several stoichiometries are well known. The MH$_6$ sodalite-like clathrate, first exemplified by CaH$_6$~\cite{RN906,RN904,RN652}, has a metal atom, M, centered in an H$_{24}$ cage consisting of six square and eight hexagonal faces. The MH$_9$ structure, which surrounds the metal atom with an H$_{29}$ cage of six irregular squares, six pentagons and six hexagons, is found in numerous rare-earth compounds including synthesized YH$_9$~\cite{RN921} and CeH$_9$~\cite{RN736,RN737,RN1015}. Further increasing the hydrogen content, MH$_{10}$ stoichiometries, where the metal atom sits within hydrogenic polyhedra of six squares and twelve hexagons, are perhaps the most well-known owing to the synthesis of the LaH$_{10}$~\cite{RN2,RN938} superhydride. The structure of the hydrogen lattice in these MH$_{10}$ systems is known as ``AST'' in the clathrate and zeolite community~\cite{doi:10.1021/jp100368e}. More recently, larger hydride cages have been predicted to encapsulate rare earth metals. These include MH$_{18}$ (M=Y, La, Ce, Ac, Th) stoichiometries with cubic ($Fmmm$ and $Fddd$) space groups, possessing H$_{36}$ cages, with a maximum predicted $T_\text{c}$ of 329~K at 350~GPa for CeH$_{18}$~\cite{doi:10.1021/jacs.2c05834}. Can we draw inspiration from these systems and other occurring clathrates, specifically with phases that contain two electropositive elements?

Looking across the periodic table to other cage-forming elements we encounter silicon, which is able to form large polyhedral structures of sp$^3$ bonded Si atoms that are filled with metal atoms, characterizing the silicon clathrates (SiCL). Two of the most well studied SiCLs are the type-I \edit{$Pm\overline{3}m$ M$_8$Si$_{24}$} (the Weaire-Phelan structure), which can be interpreted as two Si$_{20}$ and 6 Si$_{24}$ cages filled with a metal atom, and the \edit{type-II $Fd\overline{3}m$} M$_{x\leq24}$Si$_{136}$ structure, which is made of 16 Si$_{20}$ and 8 Si$_{28}$ cages with a variable number of metal atoms that can be controlled by the synthesis conditions~\cite{BEEKMAN2006111,VOLLONDAT2022163967,AMMAR2004393,RN1021,RN1020}. Several of the known silicon clathrates are superconductors, for example the type-I Ba$_8$Si$_{46}$~\cite{doi:10.1021/ic990778p}, with an ambient pressure $T_\text{c}$ of 8~K. Recent experiments report the synthesis of (Ba/La/Eu)$_4$H$_{23}$~\cite{doi:10.1021/acs.jpclett.1c00826,RN1014,doi:10.1021/acs.jpclett.0c03331} high pressure hydrides in the type-I structure, with untested $T_\text{c}$s, and Lu$_4$H$_{23}$~\cite{Li_2023}, with a $T_\text{c}$ of 73~K at 218~GPa. Furthermore, predicted Li$_2$(La/Y)H$_{17}$~\cite{PhysRevB.106.024519} both match the type-II structure, with the highest $T_\text{c}$ being 156~K at 160~GPa for Li$_2$LaH$_{17}$. These structure types have also been explored for Li$_2$NaH$_{16}$, Li$_2$NaH$_{17}$, and LiNa$_3$H$_{23}$~\cite{an2023thermodynamically}, predicting a maximum $T_\text{c}$ of 340~K at 300~GPa for Li$_2$NaH$_{17}$. Prior to these structural correlations being drawn, a computational study predicted that when the MgH$_{16}$~\cite{RN697} structure was doped with Li, an $Fd\overline{3}m$ Li$_2$MgH$_{16}$ ``hot superconductor''~\cite{RN754}, which differs from the type-II \edit{clathrate structure by a single unoccupied Wyckoff site (a clathrate atom at the (8a)~1/8,~1/8,~1/8 site)}, possessing a record calculated maximum $T_\text{c}$ of 473~K at 250 GPa could be stabilized~\cite{RN754}.

Herein, through careful DFT calculations we investigated if a high-temperature superconductor analogous to Li$_2$MgH$_{16}$~\cite{RN754}, but where Mg is replaced by Ca, could be stabilized under pressure. Because the type-I and type-II SiCL structures each possess two cages with distinct sizes, we hypothesized they would be particularly well suited templates for incorporating mixed metal atoms, and in our specific case lithium would fill the smaller cages and calcium the larger ones. \edit{Lithium is both electropositive, donating charge to the surrounding hydrogen lattice, and light, allowing for engaging in coupled high-frequency phonons with the hydrogen lattice.} Furthermore, as several Ca hydrides are synthesized or predicted superconductors, we intended to improve upon their properties with the addition of Li. With these motivations, we performed CSP searches at megabar pressures for Li$_2$CaH$_n$ ($n=10-20$). Both stable and metastable phases were discovered, and their electronic structures were analyzed. Two dynamically stable metallic structures, one that is a distortion of the type-II SiCL, and one that can be derived from it, were found, and electron-phonon coupling calculations were performed to estimate their isotropic Eliashberg $T_\text{c}$s: 330~K at 350~GPa for $Fd\overline{3}m$ Li$_2$CaH$_{16}$, and 205~K at 160~GPa or 370~K at 300~GPa for $R\overline{3}m$ Li$_2$CaH$_{17}$. This work extends the list of predicted ``hot'' superconducting ternary hydrides by two.

\section{Methods}

Exploratory crystal structure prediction (CSP) searches using the open-source XTALOPT~\cite{Lonie:2011,Falls:2019} evolutionary algorithm (EA) 12th~\cite{Avery:2019} release were performed to identify stable and metastable phases. EA runs were performed for the following stoichiometries: Li$_2$CaH$_n$ ($n$=10-20) and Li$_2$Ca at 100, 200, and 300 GPa. The initial generation consisted of random symmetric structures created by the RANDSPG algorithm.\cite{RN796} Duplicate structures were identified via the XTALCOMP algorithm for removal from the breeding pool.\cite{RN59} The breeding operators employed and their relative probabilities were: (i) stripple (50\%), (ii) permustrain (35\%), and (iii) crossover (15\%). The minimum number of generated structures was set to 1500. Each structure optimization consisted of four consecutive geometry optimizations, with each increasing in precision. The final geometry optimization step of the EA search employed a cutoff energy of 400 eV and a $k$-grid generated using the $\Gamma$-centered Monkhorst-Pack scheme~\cite{Monkhorst:1976}, with the number of divisions along each reciprocal lattice vector chosen such that the product of this number with the real lattice constant was 30~\AA{}. After the EA searches concluded, the three most enthalpically favored unique structures at each stoichiometry and pressure were used for further analysis.

Precise geometry optimizations, electronic structure calculations, electron localization functions (ELF), Bader charges, and first principles molecular dynamics (FPMD) simulations were performed using DFT calculations with the PBE functional as implemented in the Vienna \textit{Ab Initio} Simulation Package (VASP). \cite{Kresse:1996,Perdew:1996} We employed a planewave basis set with an energy cutoff of 600~eV, along with PAW potentials~\cite{Blochl:1994} where the H 1\textit{s}$^1$, Li 1\textit{s}$^2$2\textit{s}$^1$2\textit{p}$^0$ and Ca 3\textit{p}$^6$4\textit{s}$^2$3\textit{d}$^{0.01}$ electrons were treated explicitly in all calculations. $k$-meshes were generated using the $\Gamma$-centered Monkhorst-Pack scheme~\cite{Monkhorst:1976}, and the number of divisions along each reciprocal lattice vector was chosen such that the product of this number with the real lattice constant was 50~\AA{}. All FPMD simulations were performed with a 2 $\times$ 2 $\times$ 2 supercell, a single $k$-point at $\Gamma$, an $NVT$ ensemble (constant number of particles, constant volume, constant temperature), Nos\'e-Hoover thermostat~\cite{PhysRevA.31.1695}, a 0.5 femtosecond timestep, and temperatures corresponding to the predicted $T_c$s. The crystal orbital Hamilton populations (COHPs) and the negative of the COHP integrated to the Fermi level (-iCOHP) were calculated using the LOBSTER package to analyze bonding. \cite{Dronskowski:1993,https://doi.org/10.1002/jcc.24300}

To determine if the predicted structures were dynamically stable, and obtain their zero-point energy, harmonic phonons were calculated using the supercell approach. In the supercell approach, Hellmann-Feynman forces were calculated from a supercell constructed by replicating the optimized structure wherein the atoms had been displaced, and dynamical matrices were computed using the PHONOPY code. \cite{Togo:2015} Quantum ESPRESSO (QE) was used to obtain the dynamical matrix and electron-phonon coupling (EPC) parameters.\cite{Giannozzi:2009} H, Li, and Ca pseudopotentials obtained from the QE pseudopotential library were generated by the Vanderbilt ultrasoft method with a 1\textit{s}$^1$ configuration for H, a 1\textit{s}$^2$2\textit{s}$^1$2\textit{p}$^0$ configuration for Li, and a 3\textit{s}$^2$3\textit{p}$^6$4\textit{s}$^2$3\textit{d}$^0$4\textit{p}$^0$ configuration for Ca. The PBE functional was employed for all QE calculations. Plane-wave basis set cutoff energies were set to 70 Rydbergs (Ry) for all systems. The Methfessel-Paxton Brillouin-zone smearing of 0.015~Ry was applied.\cite{Methfessel:1989} The EPC parameter $\lambda$ was calculated using Gaussian broadenings in increments of 0.005 Ry ranging from 0.005 to 0.050 Ry. A 15 $\times$ 15 $\times$ 15 $k$-grid with a $q$-mesh of 5 $\times$ 5 $\times$ 5 was employed for Li$_2$CaH$_{17}$ at 300 GPa. Subsequent calculations on Li$_2$CaH$_{17}$ utilized a $q$-mesh of 3 $\times$ 3 $\times$ 3. \edit{This sparser $q$-mesh resulted in a deviation in $\lambda$ of <2\%, with a 20~K reduction in $T_\text{c}$, and an overall reduction in computational cost by 80\%}. This was determined to be sufficiently converged for prediction of EPC properties (Table S7). Therefore, a 15 $\times$ 15 $\times$ 15 $k$-grid with a \emph{q}-mesh of 3 $\times$ 3 $\times$ 3 was utilized for Li$_2$CaH$_{16}$. For Li$_2$CaH$_{17}$ the EPC appeared to be converged with these parameters at a broadening of 0.030 Ry, and Li$_2$CaH$_{16}$ converged at a broadening of 0.050 Ry. 

Critical temperature ($T_\text{c}$) values were obtained with a Coulomb repulsion parameter, $\mu^*$, of 0.1 by \edit{the Allen-Dynes modified McMillan equation for strong coupling,
\begin{equation}
T\textsubscript{c} = \frac{f_1f_2\omega\textsubscript{log}}{1.2}\exp\left[-\frac{1.04(1+\lambda)}{\lambda-
\mu^*(1+0.62\lambda)}\right], 
\label{eq:mad}
\end{equation}
where $\omega\textsubscript{log}$ is the logarithmic average frequency and the $f_1$ and $f_2$ correction factors for strong coupling and shape dependence are functions of $\omega$, $\lambda$, and $\mu^*$ (equaling unity for the weak coupling limit, see Ref.\ \cite{RN647}). $T_\text{c}$s were also obtained by solving the isotropic Eliashberg equations \cite{Eliashberg:1960} within the Fermi-surface-restricted (FSR) approximation numerically based on the spectral function, $\alpha^2F(\omega)$, obtained from the QE calculations. Isotropic solutions of $T_\text{c}$ values in hydrides have been found to underestimate $T_\text{c}$s by $\leq$ 10\% compared to anisotropic solutions~\cite{PhysRevB.81.134506,PhysRevB.101.104506,PhysRevB.104.L020511,PhysRevMaterials.6.034801}, and a recent implementation has made it possible to compare the results of the full bandwidth (FBW) approach to those obtained with the FSR approach for hydrides\cite{RN1082}. The Matsubara frequency cutoff was set to a value of 20 $\times$ $\omega_{2}$, the root-mean-square phonon frequency, while the number of frequencies was allowed to vary up to 40,000.}

\section{Results and Discussion}

\subsection{Hydrogenic Motifs}

Let us take a closer look at the predicted Li$_2$CaH$_n$ structures, to better understand the peculiarities of their atomic configurations, and in particular the hydrogenic motifs that are so important for their superconducting properties. Based on the extensive work performed on binary high-pressure hydrides containing electropositive elements~\cite{RN750,RN910}, it is known that the presence of H$^-$ and H$_3^-$ anions is detrimental for superconductivity. \edit{When these molecules are present, the hydrides are metallized via pressure-induced broadening of the H$^-$ and H$_3^-$ bonding and anti-bonding bands, resulting in a low density of states (DOS) at the Fermi level ($E_\text{F}$) and no or low $T_\text{c}$~\cite{RN654,RN660}}. Typically, compounds with H$_2^-$ motifs are predicted to possess intermediate $T_\text{c}$s owing to the metallicity that stems from the partial filling of the antibonding H$_2$ $\sigma_u^*$ orbitals, and a comparatively high DOS at $E_\text{F}$ ($N(E_\text{F})$).  \edit{Expanding from molecules into extended hydrogenic lattices, a high $N(E_\text{F})$ can be maintained. Compounds with cage-like lattices that enclathrate electropositive atoms are found to possess large values of $\lambda$, the electron-phonon coupling constant, leading to the highest computed and measured superconducting critical temperatures.\cite{RN807}}

Analysis of the structures found in our CSP searches reveals these same hydrogenic motifs, and illustrates how they vary as a function of hydrogen content and pressure (Table \ref{tab:hydrogen_motifs}). The Li$_2$CaH$_{n}$ $n=10-13$ stoichiometries were typically characterized by H$^-$ and bent, asymmetric H$_3^-$ units, interspersed with H$_2$ molecules at all pressures investigated. The most stable $n=14-15$ phases featured these same motifs at 100 and 200~GPa, however by 300~GPa incomplete clathrate cages began to emerge, suggesting that both the metal to hydrogen ratio and pressure are key for clathrate formation in the high-pressure hydrides. A clathrate cage is found within Li$_2$CaH$_{16}$ at 300~GPa, but it dissociates to an incomplete clathrate at 200~GPa and further to a hydrogenic lattice containing discrete H$_3^-$, H$_2$ and H$^-$ units by 100~GPa. The global minimum Li$_2$CaH$_{17}$ phase possesses a clathrate structure at both 300 and 200~GPa, but dissociates to form H$_2$ and H$^-$ by 100~GPa. Within Li$_2$CaH$_{18}$ and Li$_2$CaH$_{19}$ incomplete clathrates are present at 300~GPa, with H$_2$ units interspersed within the lattice for $n=18$. At pressures of 100 and 200~GPa, these incomplete clathrates dissociate to form H$_3^-$, H$_2$ and H$^-$. Lastly, Li$_2$CaH$_{20}$ possesses only H$_2$ units at 100, 200, and 300~GPa. Our observations are in-line with a chemical pressure analysis, which illustrates that clathrate formation in the high-pressure hydrides depends on the size-match between the metal cation and the hydrogenic cluster encapsulating it \cite{Hilleke:2022a}. We postulated that the Li$_2$CaH$_{16}$ and the Li$_2$CaH$_{17}$ structures could be candidates for high-temperature, and perhaps ``hot'' superconductivity, given the proclivity of clathrate-structure-types for large EPC values.

\begin{table}[ht!]
\edit{
	\caption{\edit{\textmd{Hydrogenic motifs for the most enthalpically favored phase of each stoichiometry  at 100, 200, and 300~GPa.}}} 
	\label{tab:hydrogen_motifs}
	\centering 
	\begin{tabular}{ccccc} 
	\hline
		\textbf{Stoichiometry} & \textbf{100~GPa} & \textbf{200~GPa} & \textbf{300~GPa}\\
		\hline \hline
Li$_2$CaH$_{10}$		&	H$_3^-$,H$_2$,H$^-$		&	H$_2$,H$^-$	&	H$_3^-$,H$_2$,H$^-$	\\
Li$_2$CaH$_{11}$    	&   H$_2$,H$^-$ 				&	H$_2$,H$^-$	&   H$_2$  \\
Li$_2$CaH$_{12}$		&	H$_2$,H$^-$				&	H$_2$,H$^-$	&	H$_3^-$,H$_2$,H$^-$	\\
Li$_2$CaH$_{13}$		&	H$_2$,H$^-$				&	H$_2$,H$^-$	&	H$_2$	\\
Li$_2$CaH$_{14}$		&	H$_2$,H$^-$				&	H$_2$,H$^-$	&	Incomplete Clathrate	\\
Li$_2$CaH$_{15}$		&	H$_3^-$,H$_2$,H$^-$ 		&	H$_2$,H$^-$	&	Incomplete Clathrate	\\
Li$_2$CaH$_{16}$		&	H$_3^-$,H$_2$,H$^-$		&	Incomplete Clathrate		&	Clathrate	\\
Li$_2$CaH$_{17}$		&	H$_2$,H$^-$				&	Clathrate	&	Clathrate	\\
Li$_2$CaH$_{18}$		&	H$_2$,H$^-$				&	H$_2$,H$^-$	&	Incomplete Clathrate, H$_2$ 	\\
Li$_2$CaH$_{19}$		&	H$_3^-$,H$_2$,H$^-$		&	H$_3^-$,H$_2$,H$^-$	&	Incomplete Clathrate	\\
Li$_2$CaH$_{20}$		&	H$_2$					&	H$_2$		&	H$_2$	\\
\hline
	\end{tabular} \\
	}
\end{table}

\subsection{Ternary Phase Diagrams}

\begin{figure}[h!]
\begin{center}
\includegraphics[width=5in]{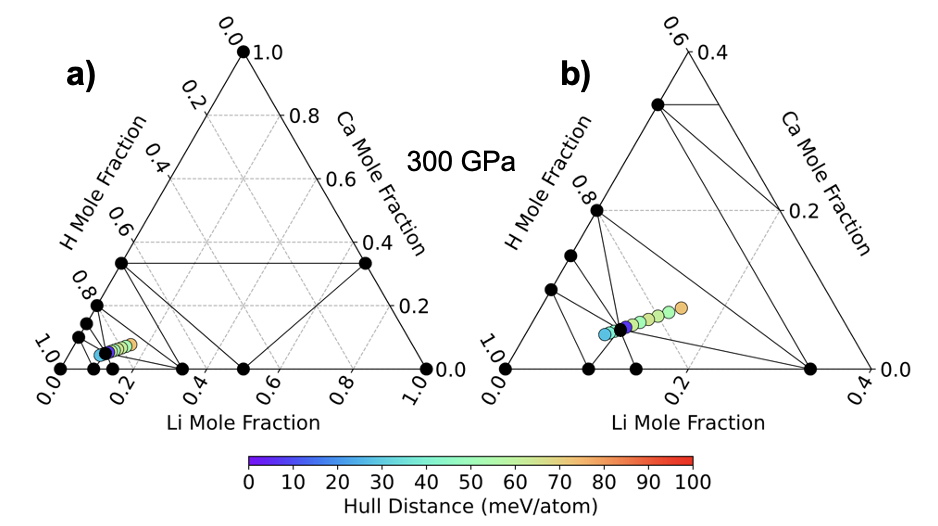}
\end{center}
\caption{\edit{Distance of the most stable Li$_2$CaH$_{n}$ ($n$=10-20) structures above the convex hull at 300~GPa. Panel (a) provides the full 3D convex hull at 300 GPa, while panel (b) focuses on hydrogen rich phases. Thermodynamically stable phases are shown in black, while metastable phases are colored based on their distance from the hull. The pure elemental phases of Ca\cite{PhysRevB.81.140106,PhysRevMaterials.2.083608}, Li\cite{PhysRevLett.106.015503}, and H\cite{RN66}, optimized to the respective pressure, were used in constructing the convex hull. For binary phases the lithium hydrides LiH, LiH$_2$, LiH$_6$\cite{RN654}, LiH$_9$, and LiH$_{10}$\cite{doi:10.1021/acs.inorgchem.6b02709} as well as the calcium hydrides CaH, CaH$_2$\cite{RN1002}, CaH$_4$, CaH$_6$\cite{RN652}, CaH$_9$\cite{RN735}, CaH$_{12}$, Ca$_2$H, and Ca$_3$H\cite{RN1002} were optimized to the respective pressure.}}
\label{fig:ternary_plot}
\end{figure}

\begin{figure}[h!]
\begin{center}
\includegraphics[width=3.33in]{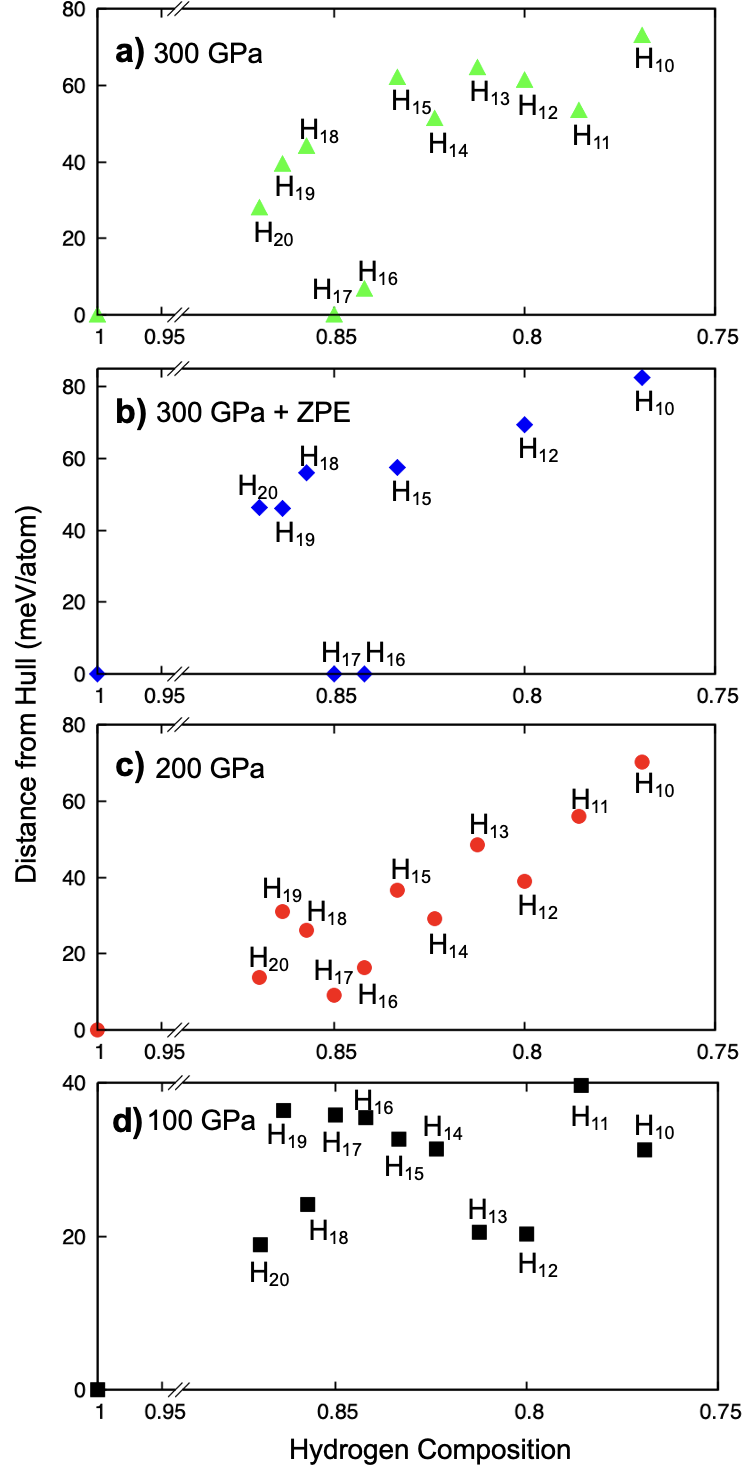}
\end{center}
\caption{\edit{Distance of the most stable Li$_2$CaH$_{n}$ ($n$=10-20) structures above the convex hull at (a,b) 300, (c) 200, and (d) 100~GPa. A two-dimensional cut of the three-dimensional Li-Ca-H convex hull corresponding to the plane between Li$_2$Ca (100 GPa) or 2Li + Ca (200 and 300 GPa) and elemental H is displayed. In panel (b) the distance is calculated using the zero-temperature enthalpies and the zero-point-energy (ZPE) corrections, and dynamically unstable phases are omitted. Li$_2$CaH$_{16}$ is included, as it was shown to be dynamically stable through first-principle molecular dynamics.}}
\label{fig:300_convex_hull}
\end{figure}

The most stable structures found in the CSP searches were employed to construct 3D convex hulls \edit{(Figures \ref{fig:ternary_plot}, S1-S3)} at 100, 200, and 300~GPa at zero-temperature and within the clamped nuclei approximation. \edit{These plots are based upon the formation enthalpy as a function of the stoichiometry. By definition the convex hull connects phases whose enthalpies are lower than that of any other phase or linear combination of phases at a particular composition. Thus, the phases that lie on the hull are thermodynamically stable, whereas those that lie above it may be important local minima provided they are close to the hull, are dynamically stable, and are protected by sufficiently high barriers. Therefore, it is often useful to plot the distance of a particular structure from the hull, to determine if it could be an important metastable phase.} Because the compositions considered all possessed a 2:1 lithium to calcium ratio while the hydrogen content varied, 2D cuts of the generated convex hulls, corresponding to a plane of increasing H content from Li$_2$Ca at 100~GPa, or 2 Li + Ca at 200 and 300~GPa were also plotted (Figure S5, S6). 
At 300~GPa a Li$_2$CaH$_{17}$ phase was predicted to be thermodynamically stable, while the structure closest to, but not on, the hull, Li$_2$CaH$_{16}$, lay 7~meV/atom above it (Table S2 and Figure \ref{fig:300_convex_hull}). Notably, as discussed above, both of these possessed hydrogenic clathrate cages.

The zero-point-energy (ZPE) is often important in determining relative energy orderings for systems containing light elements, with the greatest potential influence on H-bearing phases, as they can have the highest frequency vibrations. For a given pressure the ZPE is affected by symmetry, with lower symmetry systems generally having larger ZPEs, and the nature of the hydrogenic motifs present within the crystal, with dihydrogen molecules having a larger ZPE than hydrogens within extended lattices such as clathrate cages. Indeed, it has been shown that the ZPE can affect the identity of the on-hull structures in high-pressure polyhydrides containing electropositive metal atoms~\cite{RN663}. Obtaining the ZPE necessitates performing phonon calculations, which allows for confirmation of dynamic stability, indicated by a lack of imaginary modes.

Phonon calculations, performed in the harmonic approximation, suggested that Li$_2$CaH$_{17}$, with a hydrogenic clathrate lattice, and Li$_2$CaH$_{n}$, $n$ = (10,12,15,18-20), which did not feature clathrate structures, were dynamically stable at 300~GPa. Li$_2$CaH$_{16}$, on the other hand, possessed two modes with imaginary frequencies. These imaginary frequencies are related to the acoustic modes, with the majority of the atomic displacement being due to hydrogen. These frequencies were found to become real and positive by 350~GPa (Figure S21), indicating stabilization due to pressure. A first-principles molecular dynamics simulation at 330~K, however, showed that the hydrogenic lattice remained intact, with no molecular H units forming. This suggests one or more of the following: the phase is dynamically stable only at finite temperature; the system may be stabilized by the inclusion of anharmonicity or quantum effects; or the fineness of the grid needed to obtain real phonons is prohibitively computationally expensive. Due to the large size of the unit cell of Li$_2$CaH$_{16}$, and the associated computational expense, further calculations investigating this discrepancy and determining the lower boundary of dynamic stability were not performed, and the ZPE \edit{for this phase} was estimated neglecting the imaginary modes. Figure \ref{fig:300_convex_hull}(b) illustrates that inclusion of the ZPE lowered the enthalpy of Li$_2$CaH$_{16}$ so that it joined Li$_2$CaH$_{17}$ on the 300~GPa hull, but the remaining non-clathrate structures became destabilized, lying at least 40~meV/atom above the hull. 

Decreasing pressure moves all of the ternary hydrides progressively further from the hull. Li$_2$CaH$_{16}$ shifts up to be 16 and 35~meV/atom above it, while Li$_2$CaH$_{17}$ is calculated to be 9 and 36~meV/atom shy of thermodynamic stability at 200 and 100~GPa, respectively (Figures S1-S6). However, these compounds may still be promising at sub-300~GPa pressures, despite not laying on the hull, provided they are dynamically and thermally stable. Analysis of the zero-temperature DFT-computed enthalpies of the compounds contained in the Materials Project Database led to the estimate that $\sim$50\% of known inorganic crystalline materials are metastable, with a median metastability energy of 15~meV/atom and a 90$^\text{th}$ percentile of 67~meV/atom~\cite{RN7}. For the specific case of high-pressure hydrides, whose synthesis at high-pressures in diamond anvil cells is often catalyzed by non-equilibrium processes such as thousands-of-Kelvin degree laser heating, numerous examples of synthesized DFT-metastable compounds are known. Some examples include Ca$_2$H$_5$\cite{RN1002}, which lays 20~meV/atom above the hull at 25~GPa, and various PH$_n$ phases, which lay 30~meV/atom above the hull within the pressure range $\sim$80-225~GPa~\cite{RN655,RN637,RN658,Fu:2016}.

\subsection{Hydrogen Clathrate Structures}

Plots of the enthalpy versus pressure of the low lying structures found in our CSP searches (Figures S7, S8) showed that $Fd\overline{3}m$ Li$_2$CaH$_{16}$ became preferred (over other phases with this stoichiometry) above 270~GPa, while for Li$_2$CaH$_{17}$ the $R\overline{3}m$ structure was the most stable above 140~GPa. From the other candidates found in our EA searches, these had the lowest enthalpies up to a maximum investigated pressure of 500~GPa. The propensity for high $T_c$ in compounds with hydrogenic clathrate lattices, the observation that $Fd\overline{3}m$ Li$_2$CaH$_{16}$ and $R\overline{3}m$ Li$_2$CaH$_{17}$ both possessed such lattices with a high degree of symmetry, and the finding that both lay on the 300~GPa convex hull upon inclusion of the ZPE, prompted us to examine them in more detail. 

\begin{figure}[h!]
\begin{center}
\includegraphics[width=5.6in]{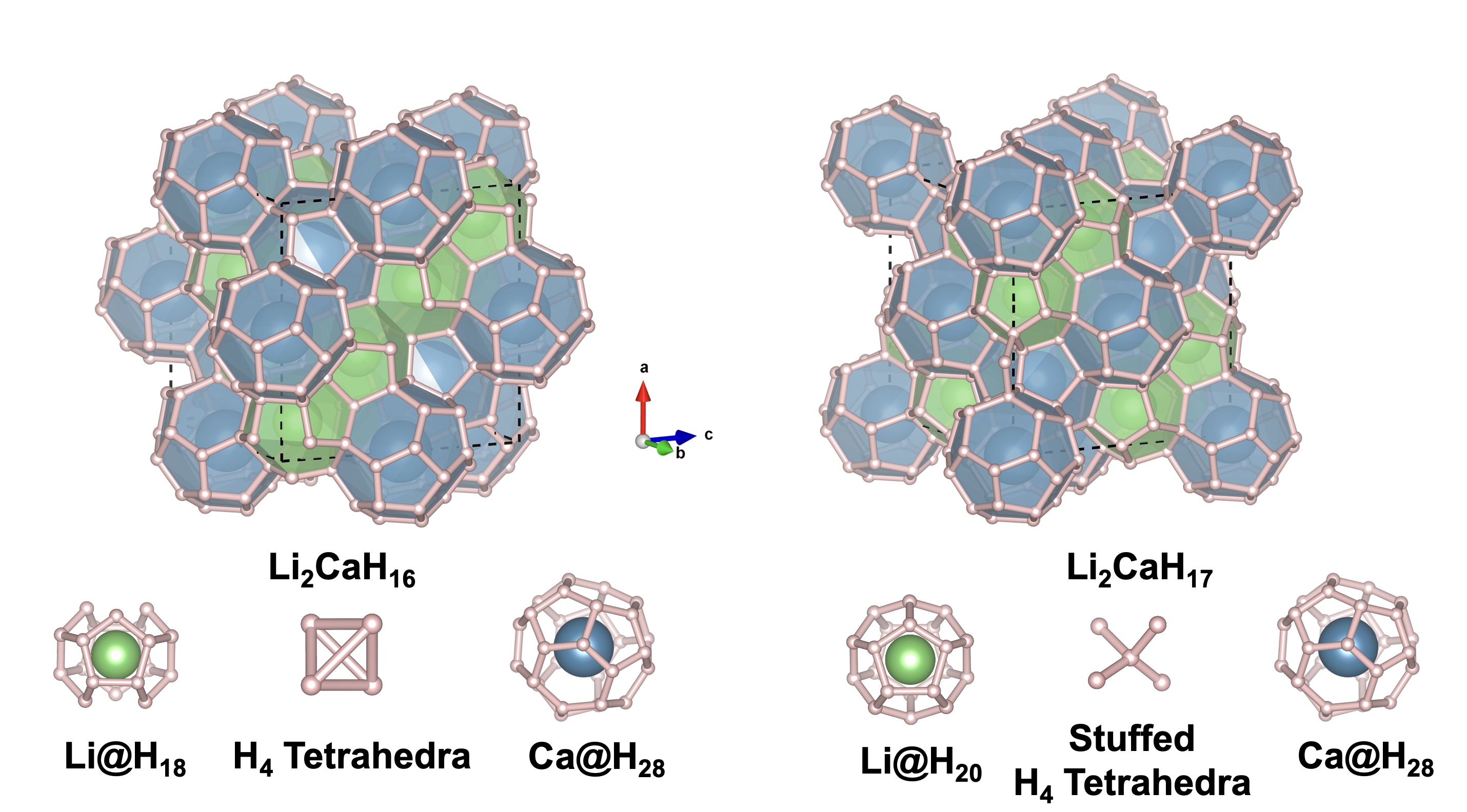}
\end{center}
\caption{\edit{Crystal structures of predicted clathrate Li$_2$CaH$_{n}$ structures. Cubic $Fd\overline{3}m$ Li$_2$CaH$_{16}$ has Li@H$_{18}$ and Ca@H$_{28}$ cages, with H$_4$ tetrahedra. Trigonal $R\overline{3}m$ Li$_2$CaH$_{17}$ (structure shown using cubic representation) has Li@H$_{20}$ and Ca@H$_{28}$ cages, with hydrogen stuffed H$_4$ tetrahedra. Calcium atoms are colored blue, lithium atoms are colored green, and hydrogen atoms are colored pink. Dashed lines represent unit cells, while the solid lines between hydrogens indicate weak bonds. Solid lines displayed for the unstuffed H$_4$ tetrahedra serve as a visual aid and are not indicative of bonding.}}
\label{fig:crystals}
\end{figure}
 
$Fd\overline{3}m$ Li$_2$CaH$_{16}$ is isotypic with previously predicted Li$_2$MgH$_{16}$ and Li$_2$NaH$_{16}$~\cite{an2023thermodynamically,RN754}. In it, each Ca atom lies in a cage surrounded by four hexagons and twelve pentagons, made of 28 hydrogen atoms, and each Li atom is enclathrated in an 18 atom hydrogen cage with four hexagons and five pentagons. The size of the cages around each metal atom are commensurate with their radii, with Ca nestled in the larger cage and lithium sitting in the smaller cage, as illustrated in Figure~\ref{fig:crystals}. At 350 GPa these polygons are highly regular, with H-H bond lengths of 1.05, 1.08, and 1.12 \r{A}. This $Fd\overline{3}m$ symmetry lattice is markedly similar to the type-II SiCL structure, differing by a single Wyckoff position. While the clearest hydrogen motifs in $Fd\overline{3}m$ Li$_2$CaH$_{16}$ are the H$_{18}$ and H$_{28}$ cages, we can also view the structure as possessing eight H$_4$ tetrahedra in the conventional cell ($Z$ = 8), corresponding to one H$_4$ tetrahedra per formula unit. These H$_4$ tetrahedra are centered around the \edit{unfilled Wyckoff position, a clathrate atom at the (8a)~1/8,~1/8,~1/8 site,} that if filled would make them isotypic to the type-II SiCL structure.
 
In $R\overline{3}m$ Li$_2$CaH$_{17}$, also shown in Figure \ref{fig:crystals}, each Ca atom is surrounded by four hexagons and twelve pentagons, and each Li atom is once again found in the smaller cage, surrounded by eleven pentagons. The additional hydrogen atom in $R\overline{3}m$ Li$_2$CaH$_{17}$ renders the $R\overline{3}m$ phase nearly isostructural with the type-II SiCLs. In comparison with $Fd\overline{3}m$ Li$_2$CaH$_{16}$, $R\overline{3}m$ Li$_2$CaH$_{17}$ possesses H$_4$ tetrahedra ``stuffed'' with the extra hydrogen atom, which assists in completing the cage around the Li ions. The stuffed hydrogen atom may stabilize this structure by balancing the chemical pressure interactions within the lattice, as has been observed in the H-stuffed H$_4$ tetrahedra of LaH$_{10}$~\cite{Hilleke:2022a}, or by allowing the structure to distort more readily. These distortions can be seen with the reduced symmetry of the faces comprising the polygons that make up the clathrate; each pentagon has H-H bond lengths between 1.08 and 1.14 \r{A}, while each hexagon has H-H bond lengths between 1.02 and 1.14 \r{A} at 300 GPa. Investigation of the $R\overline{3}m$ Li$_2$CaH$_{17}$ phase shows that these bonds become less uniform as pressure decreases. The $R\overline{3}m$ phase was determined to be dynamically stable down to 160~GPa, however the full clathrate structure does not persist to this pressure, due to the H-H bond lengths increasing (e.g.\ see ELF plot in Figure S20).

$R\overline{3}m$ Li$_2$CaH$_{17}$ was not anticipated to be the most stable spacegroup for this stoichiometry. Previous studies have predicted an $Fd\overline{3}m$ M$_3$H$_{17}$ phase, where M is a metal atom, in compounds such as Li$_2$(Sc/La/Y/Na)H$_{17}$~\cite{PhysRevB.106.024519,an2023thermodynamically}. These $Fd\overline{3}m$ symmetry M$_3$H$_{17}$ phases are isostructural with the type-II SiCL, while the $R\overline{3}m$ Li$_2$CaH$_{17}$ phase can be seen as a distortion of this structure. The external pressure, internal (or chemical pressure) and the number of valence electrons that are transferred from the metal atoms to the hydrogen cage, and potential metal-hydrogen interactions may be important factors driving this distortion. The external pressure most directly affects the $pV$ (pressure-volume) term to the enthalpy, often favoring the formation of closed packed structures at higher pressures. The chemical pressure is related to the ionic radius of the metal atom and how it fits into the cage, with ions that are too-small or too-large resulting in distorted lattices~\cite{Hilleke:2022a}. The number of valence electrons that are donated from the electropositive element to the hydrogenic lattice can influence the types of hydrogenic motifs that are present~\cite{RN652}, and the resulting H-H distances, which are also influenced by interactions such as Ca d~$\rightarrow$ H$_2$ $\sigma^*$ backbonding~\cite{RN1002}. To confirm that the $R\overline{3}m$ phase is indeed favored over $Fd\overline{3}m$, the cubic structure was built and optimized. Comparison of the 300~GPa enthalpies revealed that the $R\overline{3}m$ phase possessed a lower internal energy, $U$, but a higher $pV$ term. The bonds in the $R\overline{3}m$ phase are more distorted than the highly ordered $Fd\overline{3}m$ phase, and these distortions likely contribute to the enthalpic favorability of the $R\overline{3}m$ structure for the metal atoms studied here. Furthermore, the $Fd\overline{3}m$ phase of Li$_2$CaH$_{17}$ was determined to be both dynamically unstable at 300 GPa, and enthalpically disfavored over the pressure range of 100-500~GPa, so it was therefore not considered for additional calculations. 

\edit{A similar phenomenon of competing phases has recently been investigated in the LaH$_{10}$ system~\cite{RN1022}. While DFT calculations with classical nuclei and neglecting anharmonic effects predicted that an $Fm\overline{3}m$ phase becomes unstable with respect to distorted phases including   rhombohedral $R\overline3m$ LaH$_{10}$ below 230~GPa, the inclusion of quantum fluctuations resulted in the dynamic stability of the $Fm\overline{3}m$ phase across a wide pressure regime. Moreover, the competing phases all converged to $Fm\overline{3}m$ LaH$_{10}$ when quantum fluctuations were applied in the structural relaxation. While it is possible that quantum fluctuations could stabilize $Fd\overline{3}m$  Li$_2$CaH$_{17}$ over the $R\overline{3}m$ phase, these calculations were not performed for several reasons. First, an increased measure of non-uniformity among the H-H bonds that make up the clathrate faces is observed for FPMD simulations at 370~K, which is in alignment with the $R\bar{3}m$ phase, but not the highly symmetric $Fd\bar{3}m$ phase. Since the thermal energy provided by the thermostat is far in excess of the stabilization energy due to quantum fluctuations in hydrides, this distortion would be expected if the cubic structure was preferred. Moreover, at no point over the pressure regime did the $Fd\overline{3}m$ phase become thermodynamically preferred over the $R\overline{3}m$ phase, in contrast to the behavior observed for LaH$_{10}$. Finally, the computational expense required was seen as overly prohibitive, though recently developed machine-learning approaches may make these types of calculations possible in the future~\cite{belli2024}.}
Compared to molecular H-H bonds at the same pressure, the H-H bond lengths in both $R\overline{3}m$ Li$_2$CaH$_{17}$ and $Fd\overline{3}m$ Li$_2$CaH$_{16}$ are longer (Table \ref{tab:hh}). Bader analysis showed that the caged Li and Ca atoms donate charge to the nearby hydrogen atoms, which may covalently bond with one another. For $R\overline{3}m$ Li$_2$CaH$_{17}$ at 160~GPa, each H atom gains on average 0.16$e^-$, with Ca losing 1.05$e^-$ and Li donating a smaller amount of charge, 0.84$e^-$. When the pressure is increased to 300~GPa, each H atom in $R\overline{3}m$ Li$_2$CaH$_{17}$ gains on average 0.15$e^-$, while Ca loses 0.89$e^-$ and Li donates 0.80$e^-$. Thus, increasing pressure appears to even out the amount of charge transferred from both the formally divalent and monovalent metal atoms, perhaps because a shorter distance between the hydrogen atom and calcium can enhance the amount of backdonation between the occupied antibonding hydrogen-lattice states and the Ca d orbitals, as previously found within compressed CaH$_4$~\cite{RN1002} and Ca-S-H ternary hydrides~\cite{Zurek:2020g}. The Bader charges calculated for $Fd\overline{3}m$ Li$_2$CaH$_{16}$ at 350~GPa are similar, with H gaining 0.15$e^-$ on average, and Ca and Li donating 0.90$e^-$ and 0.78$e^-$, respectively.

\begin{table}[ht!]
	\caption{\textmd{Distances and -iCOHPs for select H-H atom pairs in $Fd\overline{3}m$ Li$_2$CaH$_{16}$, $R\overline{3}m$ Li$_2$CaH$_{17}$, $C2/c$ H$_2$, and $Im\overline3m$ CaH$_6$. The ``Edges'' column specifies how many bonds of that length make up the edges of hexagons, pentagons, and/or squares that comprise the clathrate faces.}} 
	\label{tab:COHP}
	\centering 
	\begin{tabular}{ccccc} 
	\hline \hline
		\textbf{System} & \textbf{Pressure (GPa)} & \textbf{r$_\text{H-H}$} (\r{A}) & \textbf{-iCOHP (eV/bond)$^{\dag}$} & \textbf{Edges} \\ 
         \hline  
		$Fd\overline{3}m$ Li$_2$CaH$_{16}$ & 350 & 1.05 & 2.12 (2.04) & 2x $\pentago$\\
		  &  & 1.08 & 1.67 (1.98) & 6x $\hexagon$, 2x $\pentago$ \\
		  &  & 1.12 & 1.44 (1.60) & 1x $\pentago$ \\
            \hline 
		$C2/c$ H$_2$ & 350 & 0.75 & 6.37 \\
		  &   & 1.14 & 1.16 \\
            \hline
		$R\overline{3}m$ Li$_2$CaH$_{17}$ & 300 & 1.00 & 2.60 (2.45) & 1x $\pentago$ \\
            &  & 1.03 & 2.09 (2.46) & 2x $\hexagon$, 1x $\pentago$ \\
            &  & 1.08 & 2.32 (2.00) & 2x $\pentago$ \\
		&  & 1.09 & 1.85 (1.83) & 2x $\pentago$ \\
            &  & 1.13 & 1.23 (1.65) & 6x $\hexagon$ \\
		&  & 1.14 &  1.46 (1.54) & 4x $\hexagon$, 2x $\pentago$ \\
            \hline
		$C2/c$ H$_2$ & 300 & 0.75 & 6.35 \\
		& & 1.14 & 1.13 \\
            \hline
            $Im\overline{3}m$ CaH$_6$ & 300 & 1.15 & 1.33 & 6x $\hexagon$, 4x $\Box$\\
            \hline \hline
	\end{tabular} \\
        $^{\dag}$ The -iCOHP values for a hypothetical lattice of hydrogen where no Li or Ca is present, but the H positions are unchanged are provided in parentheses.
        \label{tab:hh}
\end{table}

To further investigate the bonding within the hydrogenic lattice, the negative of the crystal overlap Hamiltonian population, integrated to the Fermi level (-iCOHP), was calculated for $Fd\overline{3}m$ Li$_2$CaH$_{16}$ and $R\overline{3}m$ Li$_2$CaH$_{17}$ at 350 and 300~GPa, respectively, and compared to the -iCOHP for the $C$2/$c$ phase of molecular hydrogen at the same pressure (Table \ref{tab:hh}). Unsurprisingly, the intermolecular -iCOHP for $C$2/$c$ H$_2$ was significantly stronger than that obtained for the shortest distance in the ternary hydrides (6.37 vs.\ 2.12~eV/bond, and 6.35 vs.\ 2.60~eV/bond, respectively). Interestingly, though the shortest intramolecular distances within $C$2/$c$ H$_2$ were nearly identical to the longest H-H bonds in $Fd\overline{3}m$ Li$_2$CaH$_{16}$ and $R\overline{3}m$ Li$_2$CaH$_{17}$, the -iCOHPs calculated for these atom pairs in the clathrate cages were somewhat stronger (1.16 vs.\ 1.44~eV/bond, and 1.13 vs.\ 1.46~eV/bond, respectively). We can compare the -iCOHPs of the Li$_2$CaH$_{16/17}$ phases with those computed for the compressed calcium polyhydrides, in particular $Im\overline{3}m$ CaH$_6$, a sodalite-like structure with a hydrogen cage. At 300~GPa, the bond measuring 1.15 \r{A} in $Im\overline{3}m$ CaH$_6$ has an -iCOHP value of 1.33~eV/bond, making it slightly weaker than the 1.14 \r{A} H-H bond in $R\overline{3}m$ Li$_2$CaH$_{17}$ at 300~GPa (1.46 eV/bond). Removing the alkali and alkaline earth metal atoms from the lattice, but without any further relaxation, yields a different set of -iCOHP values, illustrating how these electropositive elements affect the bonding in the hydrogen lattice. Though in some cases the interactions are strengthened and in other times they are weakened by the presence of the metal atoms, overall the -iCOHPs computed for the empty cages become more equalized with a somewhat smaller spread. \\

\subsection{Electronic Structure and Superconductivity}
\begin{figure}[ht!]
\begin{center}
\includegraphics[width=5in]{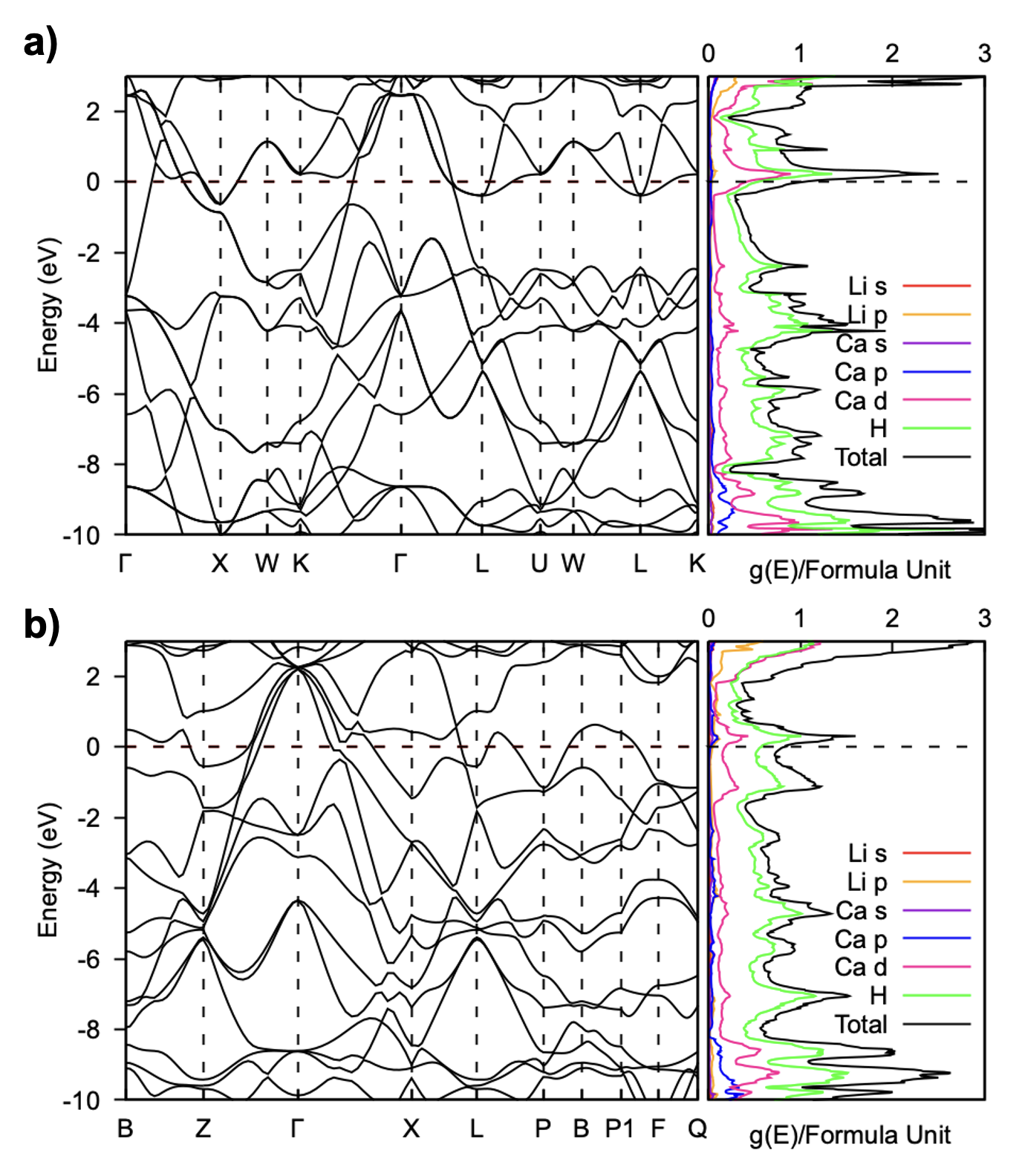}
\end{center}
\caption{The electronic band structure and projected density of states, with the Fermi energy set to 0 eV for (a) $Fd\overline{3}m$ Li$_2$CaH$_{16}$ at 350 GPa and (b) $R\overline{3}m$ Li$_2$CaH$_{17}$ at 300 GPa.}
\label{fig:dos_per_fu}
\end{figure}

$Fd\overline{3}m$ Li$_2$CaH$_{16}$ and $R\overline{3}m$ Li$_2$CaH$_{17}$ are calculated to have a high $N(E_\text{F})$ (Figure \ref{fig:dos_per_fu}), primarily of hydrogen $s$ character, boding well for superconductivity. The contribution of the calcium states to $N(E_\text{F})$ is intermediate, while the contribution of the lithium states is minimal. Therefore, the DOS plots are consistent with a Li$^+$ cation, and a not-fully-ionized calcium, resulting from the back-donation of charge from the hydrogenic lattice to the Ca d orbitals, as supported by the large amount of Ca d states at $E_\text{F}$. In $Fd\overline{3}m$ Li$_2$CaH$_{16}$ at 350~GPa and $R\overline{3}m$ Li$_2$CaH$_{17}$ at 300~GPa $E_\text{F}$ lies just below a peak in the DOS, suggesting that further electron doping (\textit{e.g.} by replacing Li via a divalent metal atom with a similar ionic radius, or Ca by a trivalent metal atom) could potentially enhance $T_\text{c}$. Such peaks in the DOS have previously been reported in $Fd\overline3m$ symmetry Li$_2$NaH$_{17}$ (just above $E_\text{F}$) and Li$_2$YH$_{17}$ (just below $E_\text{F}$) at 300~GPa~\cite{PhysRevB.106.024519,an2023thermodynamically}. Assuming a full transfer of electrons from the electropositive element to the hydrogenic lattice yields three-donated-electrons in Li$_2$NaH$_{17}$ and four-donated-electrons in Li$_2$CaH$_{17}$. In both of these systems $E_\text{F}$ sits below the peak in the DOS, so more electrons must be added. However, in Li$_2$YH$_{17}$, with five-donated-electrons, $E_\text{F}$ lies just above the peak. Thus, the optimal number of electron required to maximize $T_\text{c}$ is between four and five. We find that adding an additional 0.38 $e^-$ per FU (formula unit) ($\sim$4 $e^-$/10 FU) to Li$_2$CaH$_{16}$ at 350~GPa and an additional 0.40 $e^-$ per FU (4 $e^-$/ 10 FU) to Li$_2$CaH$_{17}$ at 300~GPa would place $E_\text{F}$ on that peak. Therefore, a quaternary compound with an approximate Li$_2$Ca$_{0.6}$M$_{0.4}$H$_{16/17}$ stoichiometry, where M is a trivalent metal ion such as Sc, Y, or La may potentially result in a higher $T_\text{c}$ than those calculated here, provided the structure is dynamically stable.

The descriptors for high-$T_\text{c}$ superconductivity -- $\lambda$ and the logarithmic average phonon frequency, $\omega_\text{log}$ -- were computed for these two phases at pressures where they were dynamically stable (Table \ref{tab:superconducting parameters}). \edit{The $\lambda$ parameter provides the strength of the electron-phonon coupling in conventional superconductors and falls in the exponential in the semi-empirical Allen-Dynes modified McMillan equation often used to estimate $T_\text{c}$ (Equation \ref{eq:mad}). The $\omega_\text{log}$ parameter lies in the pre-factor of Equation \ref{eq:mad}. A short overview of the various methods that are commonly used to estimate $T_\text{c}$ can be found in Ref.\ \cite{RN910}.}

\begin{figure}[ht!]
\begin{center}
\includegraphics[width=5in]{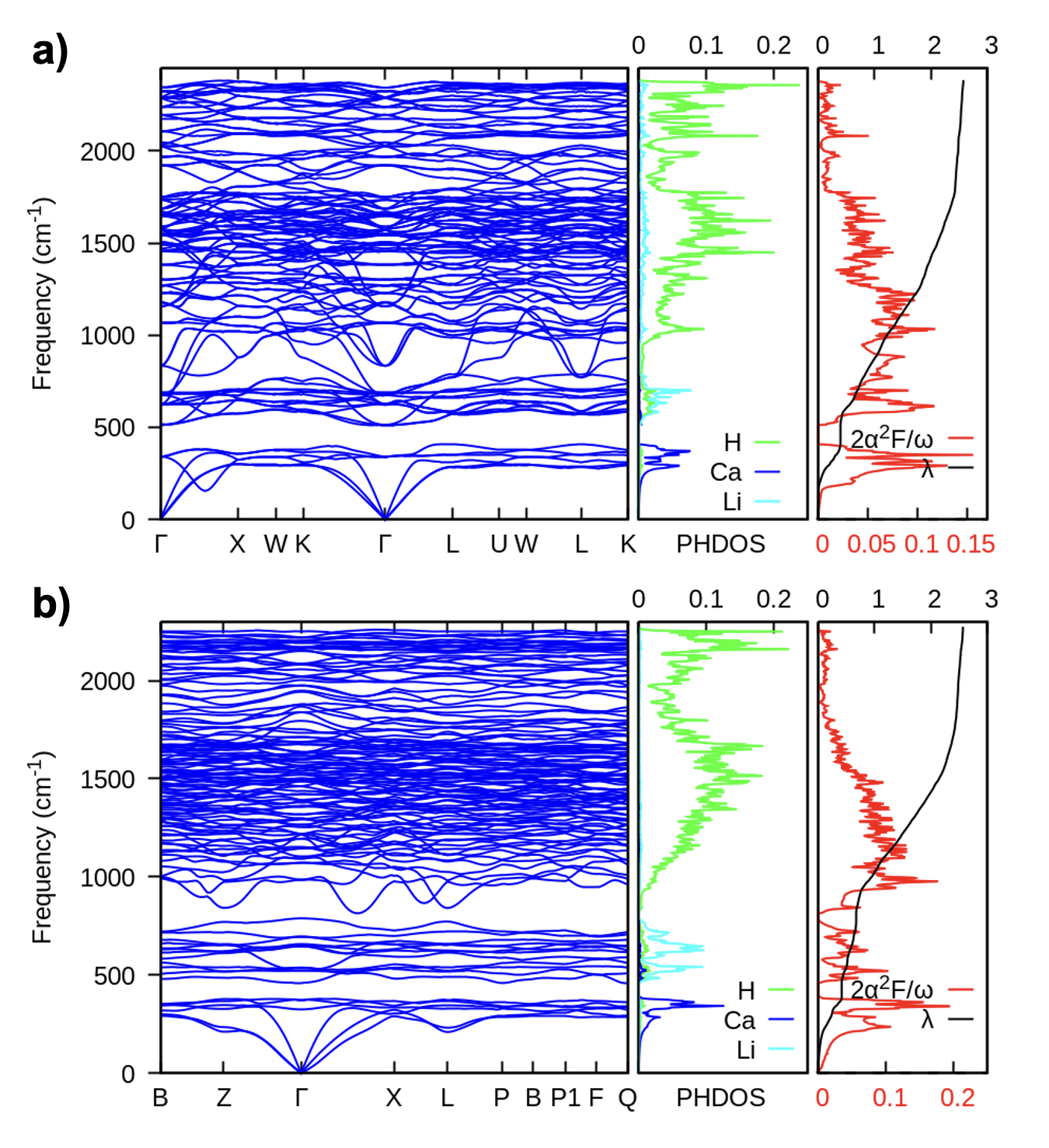}
\end{center}
\caption{Phonon dispersion curves, projected phonon density of states (PHDOS), Eliashberg spectral function scaled by the frequency ($2\alpha^2F/\omega$), and the integrated electron-phonon coupling constant ($\lambda$). (a) Properties for $Fd\overline{3}m$ Li$_2$CaH$_{16}$ at 350 GPa. (b) Properties for $R\overline{3}m$ Li$_2$CaH$_{17}$ at 300 GPa.}
\label{fig:epc}
\end{figure}

For $Fd\overline{3}m$ Li$_2$CaH$_{16}$, imaginary modes were absent only at 350~GPa, where this phase possessed the largest $N(E_\text{F})$ that we computed. The large $\lambda$ of 2.58 necessitated that the Eliashberg equations be solved numerically, and using a value of 0.1 for the Coulomb shielding parameter, $\mu^*$, we obtained a $T_\text{c}$ of 330~K for $Fd\overline{3}m$ Li$_2$CaH$_{16}$ at this pressure. Though this predicted $T_\text{c}$ is over 100~K lower than that computed for the isotypic Li$_2$MgH$_{16}$ at 250~GPa~\cite{RN754}, it surpasses room temperature, rendering Li$_2$CaH$_{16}$ a ``hot'' superconductor. Its phonon band structure, phonon projected densities of states and Eliashberg spectral function are shown in Figure \ref{fig:epc}(a). Because of the mass differences between the three atom types, the phonons can be well separated into regions where the contributions from a particular atom are dominant. Therefore, the total EPC could be partitioned into contributions originating from vibrations in the low-frequency regime, projected mostly onto the calcium atoms (15\%), followed by a lithium-dominant region (9\%), then by the mainly hydrogenic modes above 750~cm$^{-1}$ (76\%). No high-frequency modes indicative of the presence of molecular hydrogen units were observed, with the highest vibrational frequency being 2385~cm$^{-1}$, in-line with the previously computed value of 2400~cm$^{-1}$ for Li$_2$MgH$_{16}$ at 300~GPa~\cite{RN754}. The much higher $\lambda$ of Li$_2$MgH$_{16}$ at 300~GPa compared to that of Li$_2$CaH$_{16}$ at 350~GPa (3.35 vs.\ 2.58) is likely in part due to the lower pressure at which it could be stabilized, as $\lambda$ was shown to decrease with increasing pressure.\cite{RN754} The larger $\lambda$ and lighter mass of Mg as compared to Ca are both factors that result in the higher $T_\text{c}$ of Li$_2$MgH$_{16}$. Another compound that resembled $Fd\overline{3}m$ Li$_2$CaH$_{16}$, $Pm\overline3n$ LiNa$_3$H$_{23}$with a type-I SiCL structure, possessed a very similar $N(E_\text{F})$, $\lambda$ and therefore a very similar $T_\text{c}$ of 310~K at 350~GPa was predicted~\cite{an2023thermodynamically}.

\begin{table}[ht!]
\edit{
    \caption{\textmd{Superconducting parameters for $Fd\overline{3}m$ Li$_2$CaH$_{16}$ and $R\overline{3}m$ Li$_2$CaH$_{17}$. $T_\text{c}$ was calculated with $\mu^*$ = 0.1, \edit{by employing both the McMillan Allen-Dynes-modified (MAD) equation~(Equation \ref{eq:mad})~\cite{RN647}} and numerically solving the \edit{isotropic} Eliashberg (El) equations. The density of states at the Fermi level, $N$(E$_F$), is provided in units of states/eV/formula unit}, and $S$ is the superconducting figure of merit~\cite{10.1146/annurev-conmatphys-031218-013413}.} 
	\label{tab:superconducting parameters}
	\centering 
	\begin{tabular}{cccccccc} 
	\hline \hline
	\textbf{System} & \textbf{Pressure (GPa)} & \textbf{$\lambda$} & \textbf{$\omega_{\text{log}}$(K)} & \textbf{$N$($E_\text{F}$)} & \textbf{$T^\text{MAD}_\text{c}$ (K)} & \textbf{$T^\text{El}_\text{c}$ (K)} & \textbf{$S$}\\ 
            \hline
			$Fd\overline{3}m$ Li$_2$CaH$_{16}$ & 350 & 2.58 & 1123 & 1.23 & 272 & 330 & 0.94\\
            $R\overline{3}m$ Li$_2$CaH$_{17}$  & 300 & 2.56 & 1314 & 0.99 & 283 & 370 & 1.22\\
            $R\overline{3}m$ Li$_2$CaH$_{17}$  & 220 & 3.42 & 922 & 0.84 & 286 & 317 & 1.42\\
            $R\overline{3}m$ Li$_2$CaH$_{17}$  & 160 & 1.72 & 1194 & 0.74 & 179 & 205 & 1.24\\
            \hline \hline
	\end{tabular} \\
	}
\end{table}

Let us now turn to $R\overline{3}m$ Li$_2$CaH$_{17}$, whose EPC was well above the threshold requiring the use of the Eliashberg formalism in the studied pressure range: 2.56 at 300~GPa and 1.72 at 160 GPa. At 300~GPa the predicted $T_\text{c}$, 370~K, was near the boiling point of water on Earth's surface, rendering this another example of a ``hot'' superconductor. This critical temperature, while lower than that predicted for $Fd\overline{3}m$ Li$_2$MgH$_{16}$, has greatly improved upon the $T_\text{c}$ of CaH$_6$. At this pressure vibrations that fell in the primarily Ca-based low frequency region contributed 15\% to the total EPC, whereas Li-dominant vibrations contributed 4\%, and primarily H vibrations contributed 81\%. The superconducting properties of $R\overline{3}m$ Li$_2$CaH$_{17}$ can be compared directly to Li$_2$ScH$_{17}$, Li$_2$YH$_{17}$, Li$_2$LaH$_{17}$, and Li$_2$NaH$_{17}$, hydride superconductors with similar structures~\cite{PhysRevB.106.024519,an2023thermodynamically} (Table S8). Li$_2$CaH$_{17}$ and Li$_2$NaH$_{17}$ not only stand out as being better metals than Li$_2$ScH$_{17}$, Li$_2$YH$_{17}$, and Li$_2$LaH$_{17}$, but also have much larger EPCs, leading to much higher predicted $T_\text{c}$s. \edit{Calculations performed with the Allen-Dynes modified McMillan equation for strong coupling (Equation \ref{eq:mad}) yielded lower estimates of the $T_\text{c}$.}

Both the $N(E_\text{F})$ and the $T_\text{c}$ decreased with decreasing pressure for Li$_2$CaH$_{17}$, although $\lambda$ was highest at 220~GPa, where $\omega_\text{log}$ was lowest. Analysis of the structure and the ELF (Figure S19) revealed that the hydrogen clathrate is beginning to break apart into molecular units at this pressure, leading to the large increase in $\lambda$ caused by the structural instability. However, these molecular units reduce $T_\text{c}$ in-line with previously observed trends~\cite{RN910}. The formation of molecular units becomes more pronounced as pressure decreases, and by 160~GPa 50\% of the hydrogen atoms constitute H$_2$ or H$_3^-$ molecules, with the other half of the hydrogens forming 9-membered rings, decreasing $\lambda$ and the calculated $T_\text{c}$ even further. The $T_\text{c}$ for $R\overline{3}m$ Li$_2$CaH$_{17}$ at 160~GPa is calculated to be 205~K, surprisingly similar to that of CaH$_6$ ($T_\text{c}$ = 210-215~K, 160-172~GPa)~\cite{RN906,RN904,RN652}. At 160~GPa, the limit of dynamical stability for $R\overline{3}m$ Li$_2$CaH$_{17}$ (Figure S23), its superconducting figure of merit~\cite{10.1146/annurev-conmatphys-031218-013413}, $S=$~1.24, is markedly similar to that of $Im\overline3m$ H$_3$S at 150~GPa, 1.31, and somewhat larger than that of $Fd\overline3m$ Li$_2$LaH$_{17}$ at 160~GPa, 0.95.\cite{RN5,PhysRevB.106.024519} The highest phonon modes for $R\overline{3}m$ Li$_2$CaH$_{17}$ are 2275~cm$^{-1}$ at 300~GPa and 2190~cm$^{-1}$ at 160~GPa. These modes can be compared to the highest phonon modes for atomic metallic hydrogen at 500~GPa being approximately 2600~cm$^{-1}$ \cite{RN955}, with those of LaH$_{10}$ being 2300~cm$^{-1}$ at 300~GPa~\cite{RN2,RN938} and for CaH$_6$ being near 2000~cm$^{-1}$ at 150~GPa\cite{RN652}.

\section{Conclusions}
Previous theoretical calculations have predicted that ``hot superconductors'', compounds that are superconducting at room temperature and above, may be stabilized under extreme pressures~\cite{an2023thermodynamically,doi:10.1021/jacs.2c05834,RN754}. Interest in these materials stems from the fact that quantum behavior at atmospheric pressures is typically only observed at low temperatures. Herein, crystal structure prediction searches combined with density functional theory calculations predict two new hot superconductors, with critical temperatures, $T_\text{c}$s, as high as 330~K at 350~GPa for $Fd\overline{3}m$ Li$_2$CaH$_{16}$ and 370~K at 300~GPa for $R\overline{3}m$ Li$_2$CaH$_{17}$. These ternary hydride phases are predicted to lie on the 300~GPa convex hull (when zero-point-energy effects are included). Li$_2$CaH$_{17}$ was found to be dynamically stable down to 160~GPa in the harmonic approximation, where its $T_\text{c}$ was predicted to be similar to previously-synthesized H$_3$S~\cite{RN5} and CaH$_6$\cite{RN906,RN904,RN652}.

The structures of $Fd\overline{3}m$ Li$_2$CaH$_{16}$ and $R\overline{3}m$ Li$_2$CaH$_{17}$ can both be described as clathrate-like hydrogenic lattices where the heavier calcium cation occupies the larger H$_{28}$ cages, and the smaller lithium cation is situated within H$_{18}$ or H$_{20}$ encapsulations. Both of these phases are related to the Type-II clathrate structure with Li$_2$CaH$_{17}$ being a rhombohedral distortion of the $Fd\bar{3}m$ symmetry lattices found in the silicon variants, and $Fd\overline{3}m$ Li$_2$CaH$_{16}$ being derived from the ideal Type-II clathrate structure via removal of a single hydrogen atom. We postulate that the $T_\text{c}$ in these ternary hydrides could be enhanced by tuning the average valence of the enclathrated electropositive metal atoms so that the Fermi level sits on the peak in the densities of states, resulting in compositions such as Li$_2$Ca$_{0.6}$M$_{0.4}$H$_{16/17}$, where M is a trivalent metal ion.

\section{Acknowledgements}
Funding for this research was provided by the NSF award DMR-2136038 and the U.S. Department of Energy, National Nuclear Security Administration, through the Chicago-DOE Alliance Center (CDAC) under Cooperative Agreement DE-NA0003975. E.Z.\ acknowledges the U.S. Department of Energy, Office of Science, Fusion Energy Sciences funding the award entitled High Energy Density Quantum Matter under Award No. DE-SC0020340. Calculations were performed at the Center for Computational Research at SUNY Buffalo (https://hdl.handle.net/10477/79221). M.R.\ would like to thank Dr.\ Francesco Belli for discussions regarding quantum fluctuations and electron-phonon coupling.

\section{Supporting Information Available}
The Supporting Information is available free of charge on the publication website, \\ \begin{tt}http://pubs.acs.org/\end{tt}. It includes the space group at various pressures, distance above the convex hull, relative enthalpies, equation of states, electronic band structures and densities of states, electron localization functions, Bader charges, phonon density of states, computational parameters in EPC calculations, plots of first-principle molecular dynamic simulation results, superconducting properties of selected phases, and structural parameters. The optimized structures were deposited in as \textit{cif} files in the Theoretical Crystallography Open Database (TCOD IDs: 30000120-30000153).

\section*{Notes}
The authors declare no competing financial interest.


\begin{tocentry}
\begin{center}
\includegraphics[width=8.25cm]{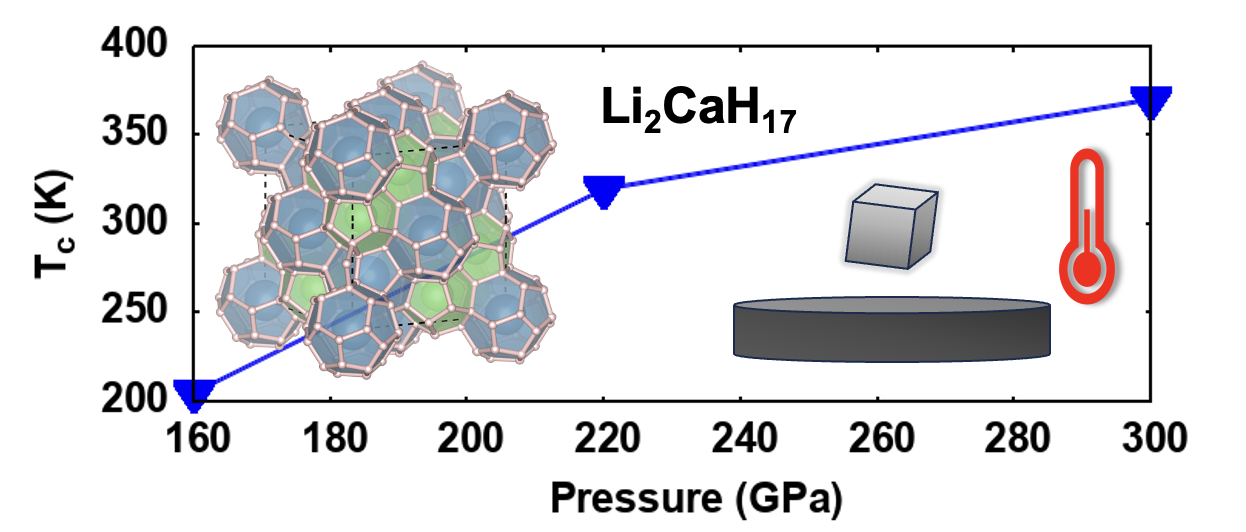}
			\label{fig:TOC}
\end{center}
\end{tocentry}



\end{document}